# Phase stability and physical properties of hypothetical $V_4SiC_3$

F. Parvin, M. M. Hossain and A. K. M. A. Islam[*]

Department of Physics, Rajshahi University, Rajshahi, Bangladesh

**Abstract**

We study the phase stability, mechanical, electronic, optical properties and Vicker's hardness of the newly predicted layered compound $V_4SiC_3$ using the first-principles method. This hypothetical compound is found to possess higher bulk modulus as well as higher hardness than those of a similar $V_4AlC_3$. The Mulliken bond population analysis indicates that the substitution of Al atom with the Si atom increases the Vicker's hardness of this compound. The electronic band structure shows that the conductivity is metallic and the main contribution comes from V $3d$ states. The partial density of states (PDOS) shows that the hybridization peak of V $3d$ and C $2s$ lies lower in energy than that of V $3d$ and Si $3p$ states which suggests that the V $3d$ - C $2s$ bond is stronger than the V $3d$ - Si $3p$ bond. The results are consistent with our bond analysis. Further we have discussed the origin of the features that appear in the optical properties. $V_4SiC_3$ is seen as a promising dielectric material showing a much better candidate material as a coating to avoid solar heating than those of $V_4AlC_3$, α-$Nb_4SiC_3$ and $Ti_4AlN_3$ compounds.

## 1. Introduction

A family of layered ternary compounds under the so called MAX phases has recently drawn a lot of interest among the scientific communities [1-22]. A comprehensive review of these activities including the synthesis of the phases in the form of bulk, films and powders has been made by Sun [1]. The $M_{n+1}AX_n$ compounds (M = transition metal, A = group A-element, mostly IIIA and IVA, X = C or N, and n = 1, 2...) have a combination of strong M-X bonds and weaker M-A bonds which possess unique and interesting combination of metallic and ceramic properties [1-3]. Although the most of the studies are on MAX phase with n =1-4, there are also a few theoretical and experimental works carried on higher order (n = 5, 6, 7) MAX phases [4 - 6]. Till now, 57 $M_2AX$ (211) compounds, 8 $M_3AX_2$ (312) compounds ($Ti_3SiC_2$, $Ti_3AlC_2$, $Ti_3GeC_2$, $Ti_3SnC_2$, $V_3SiC_2$, $(V_{0.5}Cr_{.05})_3AlC_2$, $Nb_3SiC_2$ and $Ta_3AlC_2$) and 11 $M_4AX_3$ (413) compounds ($Ti_4AlN_3$, $Ti_4SiN_3$, $Ti_4SiC_3$, $Ti_4GeC_3$, $Ti_4GaC_3$, $Ta_4AlC_3$, $Ta_4SiC_3$, $Nb_4AlC_3$, α-$Nb_4AlC_3$, $V_4AlC_3$ and $V_4AlC_{2.69}$) have been studied [1, 4, 7-16]. Among these quite a good number of phases have already been synthesized and analyzed [4, 7-14].

Among the n = 4 MAX phases, the $Ta_4AlC_3$ has been synthesized with two polymorphic crystal structures [7, 17] while $Ti_4AlN_3$, $Ti_4SiC_3$, $Ti_4GeC_3$, $Nb_4AlC_3$, $V_4AlC_3$ and $Ti_4GaC_3$ exhibit only α-type crystal structure [18, 19]. A thermodynamic study of $Ta_4AlC_3$ using first-principles method shows that α-polymorph is more stable at temperature below 1875 K, whereas β-$Ta_4AlC_3$ may exist above this critical temperature [19]. The α-polymorphs of $Ti_4AlN_3$, $Nb_4AlC_3$ and $V_4AlC_3$ show stable polymorphs at temperature up to 3000 K [1, 19]. The crystal structure of these 413 phases is hexagonal unit cell with space group $P6_3/mmc$. The stacking sequence is such that every four M-layers are separated by one A- layer. The X atoms occupy octahedral sites between the M atoms, forming a network of corner-sharing octahedra. Hu et al. [20] was able to synthesize $V_4AlC_3$ by reactive hot pressing from a V, Al and C powder mixture at 1700 °C. A theoretical study on various properties of $V_4AlC_3$ was made by Li et al. [21] who found that the lower hardness of the compound is due to the weaker Al–V covalent bond. A similar result is found in the case of synthesized $Nb_4AlC_3$ [22]. Li et al. [21, 22] predicted that the hardness of these compounds can be improved by substituting Al atoms with other atoms. In view of this and in the search of finding a better material we propose a layered-ternary compound $V_4SiC_3$ which could reveal its potential for improved behavior. For this it is important to show the compound to be thermodynamically and mechanically stable. We would first study its phase stability and then predict its elastic, electronic, optical properties and Vicker's hardness.

---

[*] **E-mail address**: azi46@ru.ac.bd (A.K.M.A. Islam)

**2. Computational methods**

The computations have been performed with the CASTEP code [23] which uses the plane-wave pseudopotential based on density functional theory (DFT). The electronic exchange-correlation energy is treated under the generalized gradient approximation (GGA) in the scheme of Perdew-Burke-Ernzerhof (PBE) [24]. The interactions between ion and electron are represented by ultrasoft Vanderbilt-type pseudopotentials for V, Si and C atoms [25]. The calculations use a plane-wave cutoff energy 450 eV for all cases. For the sampling of the Brillouin zone, $9 \times 9 \times 2$ k-point grids generated according to the Monkhorst-Pack scheme [26] are utilized. These parameters are found to be sufficient to lead to convergence of total energy and geometrical configuration. Geometry optimization is achieved using convergence thresholds of $5 \times 10^{-6}$ eV/atom for the total energy, 0.01 eV/Å for the maximum force, 0.02 GPa for the maximum stress and $5 \times 10^{-4}$ Å for maximum displacement.

**3. Results and discussion**

**3.1 Phase stability**

The newly predicted layered-ternary compound $V_4SiC_3$ is assumed to have a crystal structure similar to $V_4AlC_3$ [21] and other $M_4AX_3$ compounds. We first perform the geometry optimization as a function of normal stress by minimizing the total energy of the proposed compound. The procedure leads to a successful optimization of structure. The crystal structure is shown in Fig. 1. The geometrically optimized lattice constants of $V_4SiC_3$ are $a$ = 2.9342Å and $c$ = 21.8869 Å, whereas for $V_4AlC_3$, the corresponding parameters are 2.9199 Å and 22.7882 Å, respectively.

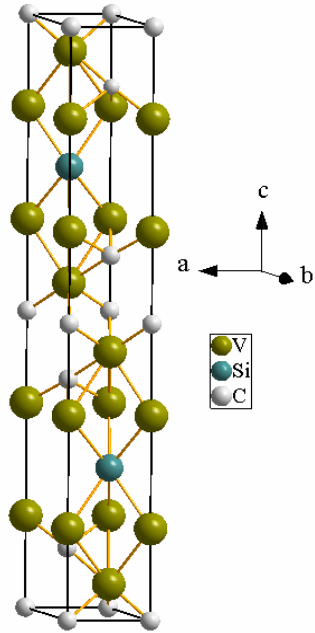

Fig. 1. The unit cell of $V_4SiC_3$.

We now investigate the phase stability of the MAX phase $V_4SiC_3$ with respect to the competing phases in the ternary phase diagram. The balance equation of $V_4SiC_3$ is as follows: $V_4SiC_3 \leftrightarrow V_3SiC_2$ + VC. The calculated energy difference, $\Delta E_{V_4SiC_3} = E_{V_3SiC_2} + E_{VC}$ is found to be -0.04 eV/f.u. It thus indicates that the phase is thermodynamically stable. It is not unlikely that the thermal stability of the MAX phases depends on their constitutive elements, the atmosphere and the vapour pressure of the elements, particularly the A-element, such as Si and Al. The phase stability at higher temperature depends on the temperature dependence of the enthalpy values and also entropy effects need to be included. Thus a more definite conclusion about the relative stabilities of the new phase at elevated temperature requires specific knowledge about both enthalpy and entropy effects.

In order to study mechanical stability we need to calculate the single crystal elastic constants of the compound. There are six elastic constants ($C_{11}$, $C_{12}$, $C_{13}$, $C_{33}$, $C_{44}$, $C_{66}$) for $V_4SiC_3$ and only five of them are independent, since $C_{66} = (C_{11} - C_{12})/2$. The calculated elastic constants $C_{ij}$ of $V_4SiC_3$ are shown in Table 1 along with other available theoretical results for comparison. All the elastic constants for $V_4SiC_3$ are positive and satisfy the well-known Born stability criteria [27]: $C_{11} > 0$, $C_{11} - C_{12} > 0$, $C_{44} > 0$, $(C_{11} + C_{12})C_{33} - 2C^2_{13} > 0$. It is seen that the material is mechanically stable under elastic strain perturbations. The value of $C_{11}$ of $V_4SiC_3$ is 43 GPa smaller than that of $V_4AlC_3$, which shows lower resistances against the principal strain $\varepsilon_{11}$. The calculated value of $C_{33}$ (14 GPa) is larger than that of $V_4AlC_3$, and this leads to higher resistances against the principal strain $\varepsilon_{33}$. $V_4SiC_3$ has nearly the same $C_{44}$ as that of $V_4AlC_3$. Therefore $V_4SiC_3$ has the same higher resistances to basal and prismatic shear deformations as $V_4AlC_3$.

Table 1. Elastic constants $C_{ij}$ (GPa).

| Phase | $C_{11}$ | $C_{12}$ | $C_{13}$ | $C_{33}$ | $C_{44}$ | $C_{66}$ |
|---|---|---|---|---|---|---|
| $V_4SiC_3$[a] | 415 | 155 | 150 | 410 | 174 | 130 |
| $V_4AlC_3$[b] | 458 | 107 | 110 | 396 | 175 | 176 |
| α-$Nb_4SiC_3$[c] | 403 | 167 | 165 | 374 | 195 | 118 |
| α-$Nb_4AlC_3$[c] | 413 | 124 | 135 | 328 | 161 | 145 |
| α-$Ta_4SiC_3$[d] | 396 | 190 | 180 | 391 | 207 | 103 |
| β-$Ta_4SiC_3$[d] | 397 | 148 | 190 | 397 | 133 | 124 |
| $Ti_4AlN_3$[e] | 420 | 73 | 70 | 380 | 128 | 173 |

[a]This work; [b][21]; [c][22]; [d][16]; [e][28].

### 3.2 Polycrystalline elastic parameters and Debye temperature

The calculated elastic constants allow us to obtain the macroscopic mechanical parameters of $V_4SiC_3$ phase, namely their bulk ($B$) and shear ($G$) moduli. These may be obtained from the set of calculated six elastic constants. According to Voigt-Reuss-Hill approximation [29, 30], the bulk modulus $B$ and shear modulus $G$ can be obtained, as listed in Table 2. The Young's modulus $E$ and Poisson's ratio $v$, are then calculated from these values using the relationships [31]: $E = 9BG/(3B + G)$ and $v = (3B - E)/6B$. We note that the bulk modulus $B$ of $V_4SiC_3$ is larger than those of $V_4AlC_3$ and α-$Nb_4AlC_3$ and this indicates to a higher resistance to volume change for the proposed compound. It is seen that the bulk modulus is greater than the shear modulus. The shear modulus $G$ is a parameter which limits the mechanical stability of this material. The value of shear modulus $G$ is smaller than that of $V_4AlC_3$ which indicates the lower resistance to shape change as $V_4AlC_3$. Again we observe that for $V_4SiC_3$ the value of $G$ is larger than those of α-$Nb_4AlC_3$ and α-$Nb_4SiC_3$, but it is smaller than that of $V_4AlC_3$.

Table 2. Bulk modulus $B$, shear modulus $G$, Young's modulus $E$ (all in GPa) Poisson's ratio $v$, anisotropic factor $A$, linear compressibility ratio $k_c/k_a$, and ratio $G/B$.

| Phase | $B$ | $G$ | $E$ | $v$ | $A$ | $k_c/k_a$ | $G/B$ |
|---|---|---|---|---|---|---|---|
| $V_4SiC_3$[a] | 239 | 146 | $E_x=333$<br>$E_z=331$<br>$E=365^*$ | 0.24 | 1.34 | 1.04 | 0.61 |
| $V_4AlC_3$[b] | 218 | 170 | $E_x=414$<br>$E_z=353$<br>$E=405^*$ |  |  | 1.21 | 0.78 |
| α-$Nb_4SiC_3$[c] | 241 | 142 | $E_x=303$<br>$E_z=278$<br>$E=356^*$ |  |  |  | 0.59 |
| α-$Nb_4AlC_3$[c] | 214 | 144 | $E_x=344$<br>$E_z=260$<br>$E=353^*$ |  |  |  | 0.67 |

[a] This work; [b] [21], [c] [22]; * Calculated by us.

Elastic anisotropic of crystals correlates with the possibility of appearance of microcracks in these materials. There are different ways to estimate elastic anisotropy theoretically. For example, the shear anisotropy factor obtained from the elastic constants $C_{ij}$, $A = 2C_{44}/(C_{11} − C_{12})$ is often used [32]. Here we see that $V_4SiC_3$ show completely anisotropic behavior. For hexagonal crystals, the ratio between linear compressibility coefficients, $k_c/k_a$ can be expressed as $k_c/k_a = (C_{11}+C_{12}-2C_{13}/(C_{33}-C_{13})$ [33]. The obtained data $k_c/k_a=1.04$ for $V_4SiC_3$ demonstrate that the compressibility along $c$-axis is greater than along $a$-axis. The ductility of a material can roughly estimated by the value of ($G/B$) [34]. It is known that if $G/B < 0.5$ the material will have a ductile behavior otherwise it should be brittle. According to this indicator (Table 1), $V_4SiC_3$ is near the borderline. Finally, the obtained value of the Poisson's ratio $v$ is 0.24. As the Poisson's ratio (υ) for the brittle covalent materials is small, and for metallic materials it is typically 0.33 [35], then the material belongs to metallic like system.

As one of the thermodynamic properties we have estimated Debye temperature $\Theta_D$ which is an important parameter which measures the vibrational response of a material. It is closely related to many physical properties such as elastic constants, specific heat and melting temperature. For temperature less than $\Theta_D$ quantum mechanical effects are very important in understanding the thermodynamic properties. The quantity is also used to estimate the electron-phonon coupling constant λ, which is proportional to the mean sound velocity $v_a$ [36]. We obtain the value of Debye temperature for $V_4SiC_3$ as 957 K. This is higher than those of α-$Ta_4SiC_3$ (535 K), β-$Ta_4SiC_3$ (502 K), $Ti_3SiC_2$ (780 K), $Ti_4AlN_3$ (762 K) and $Ti_3AlC_2$ (758 K) [16]. This indicates that the material is hard with a large wave velocity and has high thermal conductivity.

### 3.3 Electronic properties

Figs. 2(a) and (b) show the calculated band structure, and total and partial density of states (DOS) of $V_4SiC_3$, respectively. It is noted that there is no energy gap between the valence band and conduction band and there is strongly anisotropic character with lower $c$-axis energy dispersion, which can be seen from the reduced dispersion along the H-K and L-M directions. This indicates that the $V_4SiC_3$ would show anisotropic metallic conductivity. We observe that V 3$d$ electrons are mainly contributing to the DOS at the Fermi level, and should be involved in the conduction properties. It is also observed that Al electrons do not contribute significantly at the Fermi level due to a scooping effect resulting from the presence of the V 3*d* states. These results are consistent with previous report on MAX phases [37]. We observe that the lowest valence bands from -14 to -9 eV mainly come from C 2*s* states, with little contribution from the V3*d* and Si 2*p* states. The energy bands between -7.5 to 0 eV are dominated by hybridized V 3*d*, Si 3*s*/3*p* and C 2*p* states. The bonding character may be described as a mixture of covalent, ionic and, due to the *d* resonance in the vicinity of the Fermi level, metallic. The PDOS shows an interesting feature: the hybridization peak of V 3*d* and C 2*s* lies lower in energy than that of V 3*d* and Si 3*p* states. This suggests that the V 3*d* – C 2*s* bond is stronger than the V 3*d* – Si 3*p* bond.

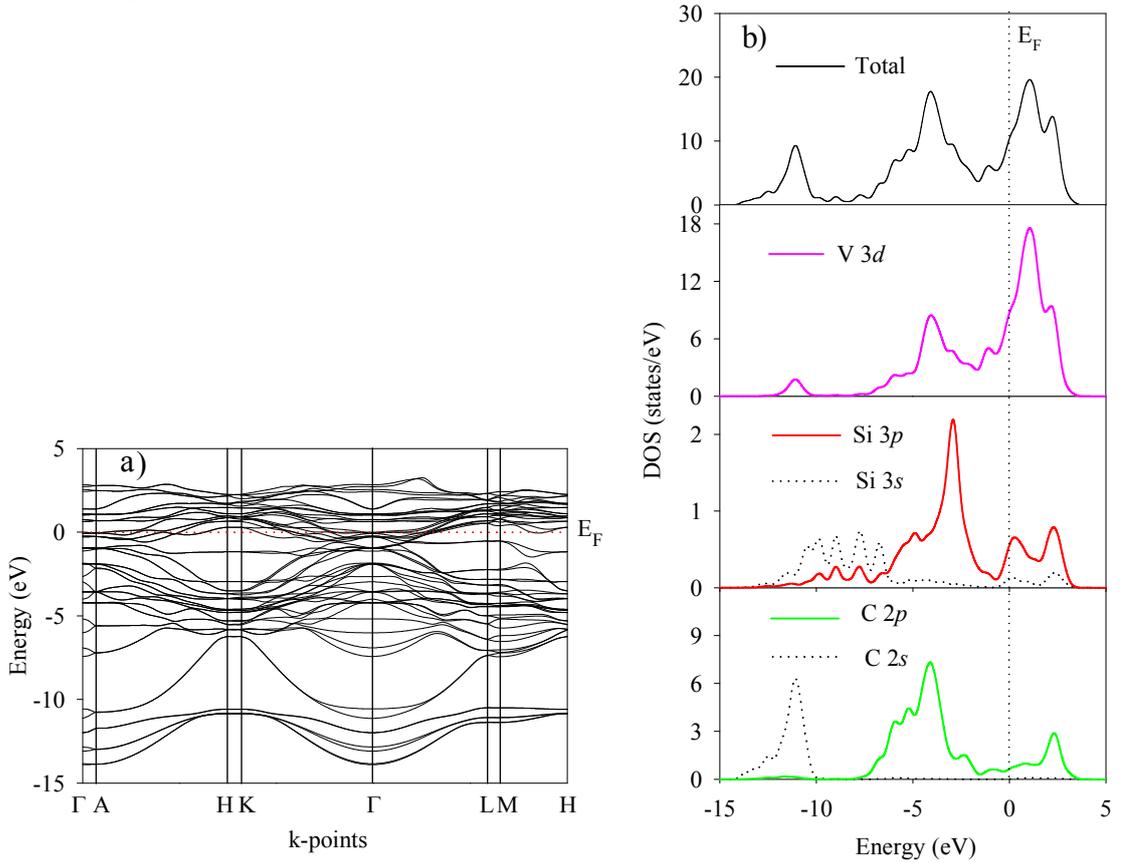

Fig. 2. (a) The electronic band structure, and (b) the total and partial density of states (PDOS) of $V_4SiC_3$.

### 3.4 Optical properties

The study of the optical functions helps to give a better understanding of the electronic structure. Fig. 4 shows the optical functions of $V_4SiC_3$ calculated for photon energies up to 20 eV along with the results of $V_4AlC_3$. We have used a 0.5 eV Gaussain smearing for all calculations. This smears out the Fermi level, so that k-points will be more effective on the Fermi surface.

Fig. 3 shows the reflectivity spectra of $V_4SiC_3$ along with the spectra of $\alpha$-$Nb_4SiC_3$, $V_4AlC_3$ and $Ti_4AlN_3$. We see that the reflectivity of $V_4SiC_3$ is a function of photon energy. We notice that the reflectivity value of $V_4SiC_3$ is always higher than those of $Ti_4AlN_3$, $V_4AlC_3$ and $\alpha$-$Nb_4SiC_3$. Therefore, the capability of the predicted $V_4SiC_3$ to reflect solar light is stronger than the other existing $Ti_4AlN_3$, $V_4AlC_3$, and $\alpha$-$Nb_4SiC_3$. We can see that the spectrum of $V_4SiC_3$ is roughly nonselective. This kind of nonselective characteristic implies that solar heating can be reduced, and the equilibrium temperature of material surface will be moderate in strong sunlight. So we conclude that $V_4SiC_3$ is also candidate material as a coating to avoid solar heating.

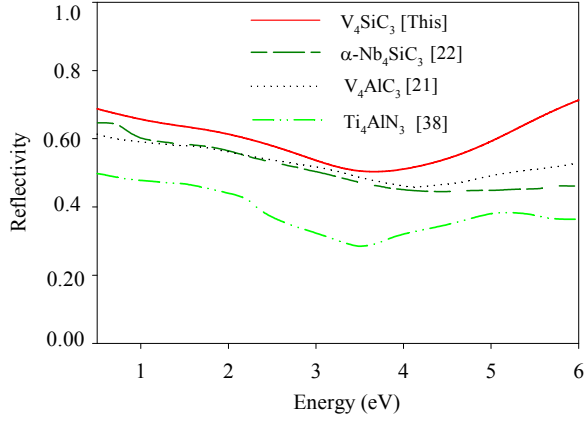

Fig. 3. Reflectivity spectra of $V_4SiC_3$, $\alpha$-$Nb_4SiC_3$, $V_4AlC_3$ and $Ti_4AlN_3$.

Fig. 4 (a) shows the dielectric function for $V_4SiC_3$ in comparison to that of synthesized $V_4AlC_3$. It is observed that the real part $\varepsilon_1$ vanishes at about 11 eV. The calculated static dielectric constant $\varepsilon_1(0)$ for $V_4SiC_3$ is 160, which is larger than those of $BaTiO_3$ (5.12), $BiInO_3$ (6.75), $Ti_3N_4$ (18.31) [39-42] and $V_4AlC_3$ (126) [21] which indicates that the material is a promising dielectric material. In the range of $\varepsilon_1 < 0$, $V_4SiC_3$ exhibits metallic reflectance characteristics. The peak of the imaginary part of the dielectric function is related to the electron excitation. For the imaginary part of $\varepsilon_2$, the peak ~ 0.25 eV is due to transitions within the V 3$d$ bands.

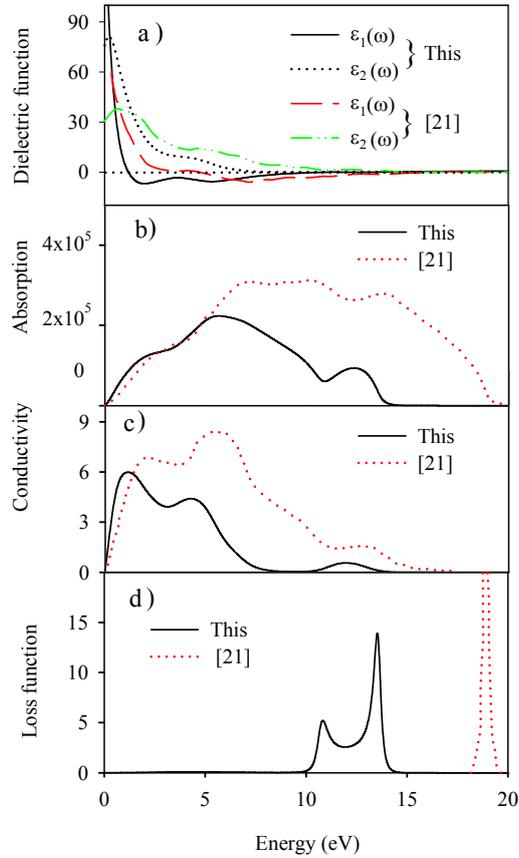

Fig. 4. The calculated optical functions of (a) dielectric function (b) absorption (c) conductivity and (d) loss function of $V_4SiC_3$.

In Fig. 4 (b), the $V_4SiC_3$ has an absorption band in the low energy range due to its metallic nature. Its absorption spectrum rises sharply and has two peaks at ~ 5.6 and ~ 12.3 eV. It then decreases to zero at ~15 eV. The origin of peaks arises to the interband transition from the Si/C $p$ to the V $d$ states. From Fig. 4 (c) we see that the photoconductivity starts when the photon energy is zero, indicating that the materials should have no band gap. Indeed no band gap is seen in the band structure calculations (Fig. 2). Moreover, the photoconductivity and hence the electrical conductivity of a material increases as a result of absorbing photons [43]. The electron energy loss function is an important optical parameter. It describes the energy loss of a fast electron traversing a material is usually large at the plasma frequency. Prominent peak is found at 11 eV, which corresponds to a rapid reduction in the reflectance. The peak also represents the characteristic associated with the plasma resonance and the corresponding frequency is the so-called plasma frequency.

### 3.5 Mulliken bonding population and Vicker's hardness

The Mulliken bond populations are calculated to understand the bonding behavior as well as to obtain Vicker's hardness ($H_V$) of the proposed compound. The relevant formula for this [44-45]:

$$H_V = \left[\prod^{\mu}\left\{740(P^{\mu} - P^{\mu'})(v_b^{\mu})^{-5/3}\right\}^{n^{\mu}}\right]^{1/\sum n^{\mu}}$$

where $P^{\mu}$ is the Mulliken population of the $\mu$-type bond, $P^{\mu'} = n_{free}/V$ is the metallic population and $v_b^{\mu}$ is the bond volume of $\mu$-type bond.

The calculated results are given in Table 3. The Mulliken bond populations explain the overlap degree of the electron clouds of two bonding atoms. Hence the highest value indicates the strong covalency of the chemical bonding. Comparing the bond overlap population, $P^{\mu}$ of $V_4SiC_3$ and $V_4AlC_3$ we found that the value of $P^{\mu}$ for Si-V bond is about 4 times larger than that of Al-V bond. In other words, it can be said that Si-V bond is about 4 times stronger than Al-V bond. Up to date, among the MAX phases $Ti_2SC$ is the hardest material with measured $H_V$= 8 GPa [1]. The theoretical hardness of $V_4SiC_3$ is found to be 12.87 GPa which is larger than that of $V_4AlC_3$. The present calculation $H_{Vcal}$ is also higher than the theoretically reported value of 10.86 GPa for α-$Nb_4SiC_3$ [22] and is the largest value obtained so far. Thus the substitution of Al atom with Si atom has improved the hardness of the proposed compound. It is observed that the Si-V bond in our compound is weaker than C-V bond, which is also consistent with our DOS calculation. The success of $Ti_2SC$ as a free-cutting element in steels compared to other known MAX phases, the largest hardness found for the proposed material will make it more suitable as a free-cutting phase to replace the harmful Pb on the environment in free-cutting stainless steels and even in most brasses [46-47].

Table 3. Calculated Mulliken bond overlap population of $\mu$-type bond $P^{\mu}$, bondlength $d^{\mu}$, metallic population $P^{\mu'}$, bond volume $v_b^{\mu}$ (Å$^3$) and Vickers hardness of $\mu$-type bond $H_v^{\mu}$ and $H_v$ of $V_4SiC_3$.

| Compounds | Species | Bond | $d^{\mu}$ (Å) | $P^{\mu}$ | $P^{\mu'}$ | $v_b^{\mu}$ (Å$^3$) | $H_v^{\mu}$ (GPa) | $H_v$ (GPa) |
|---|---|---|---|---|---|---|---|---|
| $V_4SiC_3$[a] | C | C-V | 2.00 | 0.96 | 0.03 | 7.53 | 23.77 | |
| | Si | | 2.08 | 0.89 | 0.03 | 8.39 | 18.38 | 12.87 |
| | V | | 2.09 | 0.93 | 0.03 | 8.57 | 18.56 | |
| | | Si-V | 2.59 | 0.50 | 0.03 | 16.33 | 3.31 | |
| $V_4AlC_3$[b] | C | C-V | 1.99 | 0.97 | 0.012 | 7.04 | 27.41 | |
| | Al | | 2.08 | 0.89 | 0.012 | 8.04 | 20.13 | 9.33 |
| | V | | 2.09 | 0.94 | 0.012 | 8.15 | 20.8 | |
| | | Al-V | 2.76 | 0.13 | 0.012 | 18.77 | 0.66 | |

[a] This work, [b] Ref. [21].

**4. Conclusions**

The predicted layered compound $V_4SiC_3$ is shown to be thermodynamically and mechanically stable under elastic strain perturbations. The Mulliken bond population analysis shows that the stronger Si-V bond is responsible for the improved Vicker's hardness of this compound. $V_4SiC_3$ possess higher resistance to volume change than that of $V_4AlC_3$ and has a slight anisotropy on elasticity. It is seen that $V_4SiC_3$ is nearly at the borderline of brittleness but it is more ductile than $V_4AlC_3$. The electronic band structure shows metallic conductivity and the main contribution comes from V $3d$ states. Again Si-V bond is weaker than C-V bond indicating that the C-V bond is more resistant to deformation than the Si-V bond. The weaker Si-V bond in $V_4SiC_3$ results in the lower $G$ and $G/B$ ratio, which indicates the high fracture toughness compared to $V_4AlC_3$. The analysis of the spectrum of dielectric function shows that $V_4SiC_3$ is a promising dielectric material than $V_4AlC_3$. It is also a good reflector and poor absorber of sunlight and has the potential to be used as a coating to avoid solar heating than the existing $V_4AlC_3$, α-$Nb_4SiC_3$ and $Ti_4AlN_3$ compounds. We expect that our predictions will inspire experimental investigations into the possible preparation by techniques such as the thin-film deposition and the properties thereof.